\documentclass{elsart41}
\usepackage{graphics}
\usepackage{graphicx}
\usepackage{amssymb}
\begin{document}

\begin{frontmatter}
%
%
%
\title{From the spin-fermion model to anisotropic superconductivity}
%
%
\author[AA]{Lizardo H. C. M. Nunes\corauthref{lizardo}},
\ead{lizardo@if.uff.br}
\author[AA]{Eduardo C. Marino}
\address[AA]{Instituto de F\'{\i}sica, Universidade Federal do Rio de Janeiro,
Caixa Postal 68528, Rio de Janeiro, 21941-972, Brazil}
\corauth[lizardo]{Corresponding author. Tel: +55 21 2562 7913
fax: +55 21 2562-7368}
\begin{abstract}

We use the spin-fermion model
to describe the CuO$_2$ planes
of the high-$T_{ c } $ superconductors. Using a large wavelength approach,
we show that the ferromagnetic component of the Cu spin fluctuations
couple to the oxygen holes producing a pairing interaction that leads to a
superconducting gap whose symmetry is determined by the anisotropy of
the Kondo interaction.
We calculate $ T_{ c } $ as a function of the hole concentration
in a mean-field approximation and our numerical results aare in good agreement
with the experiments.

\end{abstract}
%

\end{frontmatter}


The spin-fermion model is very useful in the description of a variety
of strongly correlated electronic systems, ranging from heavy-fermions
to manganites and high-$ T_{ c } $ superconductors.
Our purpose in this note is to show that, integrating over the spin degrees
of freedom, in the large wavelength limit of this model, we obtain an effective interaction for the
electrons that produces a $ d $-wave symmetric superconducting gap. We numerically solve the gap equation
and display the superconducting transition temperature Tc as a function of the occupation number.
A crucial ingredient for this result, as we shall see, is the anisotropy of the Kondo interaction.

We start from the spin-fermion Hamiltonian
\cite{Kampf94} envisaging the description of
the CuO$_{ 2 } $ planes of the cuprates,
\begin{eqnarray}
H
& = &
- t_{ pp } \sum_{ \langle { \bf  j j' } \rangle, \sigma }
c^{ \dagger }_{ { \bf j}, \sigma } c_{ { \bf j' }, \sigma }
+
J_{ H }
\sum_{ \langle { \bf  i, i' } \rangle }
{\bf S }_{\bf i }
\cdot
{\bf S }_{ {\bf i' } }
\nonumber
\\
& &
+
\sum_{ \langle { \bf  i, j, j' } \rangle }
J^{ {\bf i, j, j' } }_K
{\bf S }_{\bf i }
\cdot
{\bf s }_{ {\bf j, j' } }
\, ,
\label{EqHspin-fermion}
\end{eqnarray}
where
$
{\bf s }_{ {\bf j, j' } } =
\sum_{ \alpha, \, \beta}
c^{ \dagger }_{ { \bf j }, \alpha }
{\bf \sigma}_{ \alpha \, \beta }
c_{ { \bf j }', \beta }
$
and $ c^\dagger_{ {\bf j }, \sigma } $
creates a hole
with spin $ \sigma $
in the oxygen $ p_{ x, y }$-orbitals.
The site indices $ {\bf i } $ and $ {\bf j } $
denote the Cu and O sites respectively and
the last term in the above Eq. (\ref{EqHspin-fermion})
is a non-local Kondo like interaction
between the spin on the Cu site, which is represented by $ {\bf S } $,
and the holes on the four surrounding O sites.

%

We shall now use the long-wavelength continuum field theory
associated to Eq. (\ref{EqHspin-fermion}). We make
$ c_{ { \bf j }, \, \sigma } \rightarrow \psi_{ \sigma } ( \bf x ) $
and using spin coherent states, we have
$ {\bf S } \rightarrow S {\bf N }( {\bf x } ) $, which are the
continuum eigenvalues of the spin in these states.

The grand-partition function
becomes
\begin{equation}
Z
 =
\int
\mathcal{ D } \psi
\mathcal{ D } \psi^{ \dagger }
\mathcal{ D } {\bf L }
\mathcal{ D } {\bf n }
e^{
i S_{ B }
-
\int_{ 0 }^{ \beta }
d \tau
\left(
\int
d^2 x
\mathcal{ L }_0
+
H_K + H_H
\right)
}
,
\label{Eqpartition}
\end{equation}
where
$ S_{ B } $ is the Berry's phase and ${\bf n}$ and ${\bf L}$ are, respectively, the
antiferromagnetic and ferromagnetic fluctuations of ${\bf N }$.
\begin{equation}
\mathcal{ L }_{ 0 }
=
\sum_{ \sigma }
\psi^{ \dagger }_{ \sigma } ( {\bf x } )
\left(
\partial_{ \tau}
+
\frac{ {\nabla }^2 }{ 2 m^* } - \mu
\right)
\psi_{ \sigma } ( {\bf x })
\, ,
\label{EqLo}
\end{equation}
is related to the hopping term of the Hamiltonian,
\begin{equation}
H_H =
\frac{ 1 }{ 2 }
\int d^{2} x
\,
\left(
\rho_s
| \nabla {\bf n } |^2
+
\chi_{ \bot } S^2
| {\bf L } |^2
\right)
\,
\label{EqHH}
\end{equation}
is the well-known
Heisenberg Hamiltonian
in the continuum limit \cite{Tsvelik},
and
\begin{eqnarray}
H_K
& = &
\frac{ S }{ a^d }
\int d^{2} x
d^2 y_1
d^2 y_2
J\left( {\bf x} ; {\bf y}_1; {\bf y}_2 \right)
\nonumber
\\
& &
\; \; \; \; \;
\times
\sum_{ \alpha, \, \beta }
\psi^{ \dagger }_{ \alpha } ( {\bf x } )
\left(
{\bf L }
\cdot
{\bf \sigma }
\right)_{ \alpha, \, \beta }
\psi_{ \beta } ( {\bf x })
\, ,
\label{EqHK}
\end{eqnarray}
is the Kondo part of the Hamiltonian,
where we have neglected oscillating terms corresponding to the antiferromagnetic fluctuations.
Moreover, the Kondo coupling in Eq. (\ref{EqHK}) is given by
\begin{eqnarray}
J_{ K }\left( {\bf x} ; {\bf y}_1; {\bf y}_2 \right)
& = &
J_K
\sum_{ {\bf a}_i , { \bf a}_j }
\eta_{ { \bf a}_i {\bf a}_j }
\delta\left[{ {\bf y}_1 - \left( {\bf x } + {\bf a}_i \right) } \right]
\nonumber
\\
& &
\ \ \ \
\times
\delta\left[{ {\bf y}_2 - \left( {\bf x } + {\bf a}_j \right) } \right]
\, ,
\label{EqJK}
\end{eqnarray}
where the coefficients
$ \eta_{ { \bf a}_i {\bf a}_j } = \pm 1 $ reflect the anisotropy of the
copper and oxygen wave functions.
 and the vector
 $ { \bf a }_{ i } $
runs over the four first-neighbors oxygen sites around a copper atom.

Integrating over $ {\bf L } $,
we obtain
$$
{\bf s }\;^2
=
- 3 \sum_{ \alpha, \beta}
\sum_{ {\bf a}_i , { \bf a}_j
{\bf b}_i , { \bf b}_j }
\eta_{ {\bf a}_i  { \bf a}_j }
\eta_{ {\bf b}_i  { \bf b}_j }\times
$$
$$
\psi^\dagger_{ \alpha }
( { \bf x } +  { \bf a}_i )
\psi^\dagger_{ \beta }
( { \bf x } +  { \bf b}_i ))
\psi_{ \beta }
( { \bf x } +  { \bf a}_j )
\psi_{ \alpha }
( { \bf x } +  { \bf b}_j ))
$$
\begin{equation}
+
4\sum_{ \sigma }
\sum_{ {\bf a}_i , { \bf a}_j }
\eta_{ {\bf a}_i  { \bf a}_j }
\psi^{ \dagger }_{ \sigma }
( { \bf x } +  { \bf a}_i )
\psi_{ \sigma }
( { \bf x } +  { \bf a}_j )
\equiv \mathcal{ L }^{ 1 }_{ \mbox{\scriptsize{eff}} }
\, ,
\label{EqLeff1}
\end{equation}
in addition to the usual dynamical term for ${\bf n }$
and a crossed term, which is eliminated by a canonical
transformation  \cite{Marino02}.

We arrive at the effective Lagrangian
\begin{equation}
\mathcal{ L }_{ \mbox{\scriptsize{eff}} }
=
\mathcal{ L }_{ 0 }
+
\sum_{ \sigma } \phi_\sigma
\psi^{ \dagger }_{ \sigma }
\psi_{ \sigma }
+
\mathcal{ L }^{ 1 }_{ \mbox{\scriptsize{eff}} }
+
\mathcal{ L }_{ \mbox{\scriptsize{NL$ \sigma $}} }
\label{EqLeff}
\end{equation}
where
$ \phi_{\uparrow(\downarrow)} =
\sum_{i = 1, 2}
+(-) z_{ i }^{ * } \partial_{ \tau } z_{ i }
+
\nabla z_{ i }^{ * }
\nabla z_{ i }
$
and
$ \mathcal{ L }_{ \mbox{\scriptsize{NL$ \sigma $}} } $
is the well-known non-linear sigma model Lagrangian.
Eq. (\ref{EqLeff}) potentially describes both
the magnetic and superconducting orderings.

Using the CP$^1$ representation for the localized spin degrees of freedom
and integrating over the CP$^1$ $z_i$-fields in the large wavelength (small ${\bf k}$)
regime, assuming $|z_i| \simeq constant$, we get
the final electronic Lagrangian
\begin{equation}
\mathcal{ L }_{ \psi }
=
\mathcal{ L }_{ 0 }
+
\mathcal{ L }^{ (1) }_{ \mbox{\scriptsize{eff}} }
+
\mathcal{ L }^{ (2) }_{ \mbox{\scriptsize{eff}} }
\label{EqLeffResult}
\end{equation}
where
\begin{equation}
\mathcal{ L }^{ (2) }_{ \mbox{\scriptsize{eff}} }
=
\frac{ \rho }{ 4 }
\sum_{ \sigma }
\psi^{ \dagger }_{ \sigma }
\psi^{ \dagger }_{ -\sigma }
\psi_{ -\sigma }
\psi_{ -\sigma }
+
\frac{ \rho }{ 4 }
\sum_{ \sigma }
\psi^{ \dagger }_{ \sigma }
\psi_{ \sigma }
\, .
\end{equation}
Notice that the first term above is a BCS interaction,
which produces $ s $-wave isotropic superconductivity
for a constant Heisenberg interaction $ J_{ H } $.

Estimated values for the coupling constants in the case of the cuprates indicate that
we may neglect the s-wave (BCS) term. Then,
using a phenomenological choice for the coefficients
$\eta_{{\bf a}_i,{\bf b}_j}$, namely
{\bf Escrever isto em uma linha...}
\begin{eqnarray}
\eta_{ {\bf a}_x,  -{ \bf a }_x } = 1
& &
\eta_{ -{\bf a}_x,  {\bf a}_x } = -1
\nonumber
\\
\eta_{ {\bf a}_y,  -{\bf a}_y } = -1
& &
\eta_{ -{\bf a}_y,  {\bf a}_y } = 1
\nonumber
\\
\eta_{ \pm {\bf a}_x,  \pm {\bf a}_y } = - \eta_{ \mp {\bf a}_y,  \mp {\bf a}_x } = 1
& &
\eta_{ {\bf a}_x,  {\bf a}_x }
=
- \eta_{ -{\bf a}_x,  -{\bf a}_x }
= 1
\nonumber
\\
\eta_{ {\bf a}_y,  {\bf a}_y }
& = &
- \eta_{ -{\bf a}_y,  -{\bf a}_y }
= 1
\, ,
\label{EqEtas}
\end{eqnarray}
we
Fourier transform the final fermionic Hamiltonian, obtaining
$$
H_{ \mbox{\scriptsize{SC}}  }
=
\sum_{ \sigma }
\int d^2 k
\left( \epsilon_{\bf  k }
- \mu
+
\frac{ \rho_{ s } }{ 4 } \right)
\psi^{ \dagger }_{ \sigma } ( {\bf k } )
\psi_{ \sigma } ( {\bf k } )
$$
\begin{equation}
-
\int d^2 k \, d^2 k'
\,
g_{ { \bf k },  {\bf k' } }
\psi^{ \dagger }_{ \downarrow }
( -{ \bf k } )
\psi^{ \dagger }_{ \uparrow }
( { \bf k } )
\psi_{ \uparrow }
( { \bf k }'  )
\psi_{ \downarrow }
( -{ \bf k}' )
\, ,
\label{EqHSC}
\end{equation}
where
$
g_{ {\bf k }, { \bf k' } }
=
\frac{ 3 J^2 }{ 2 \chi_{ \bot } }
{ \eta }_{ \bf k }
{ \eta}_{ \bf k' }
$,
with
$
\eta_{ \bf k } =
\sin \left(   k_x a\right) - \sin\left(  k_y a\right)
$, which produces a superconducting gap with line nodes,
$\Delta({\bf k})= \Delta_0 \ \eta_{ \bf k }$.

From Eq. (\ref{EqHSC}),
we calculate $ T_{ c } $
as a function of the hole occupation in a mean-field approximation,
as in the standard BCS approach,
and the chemical potential is calculated self-consistently.

\begin{figure}[!ht]
\begin{center}
\includegraphics[width=0.45\textwidth]{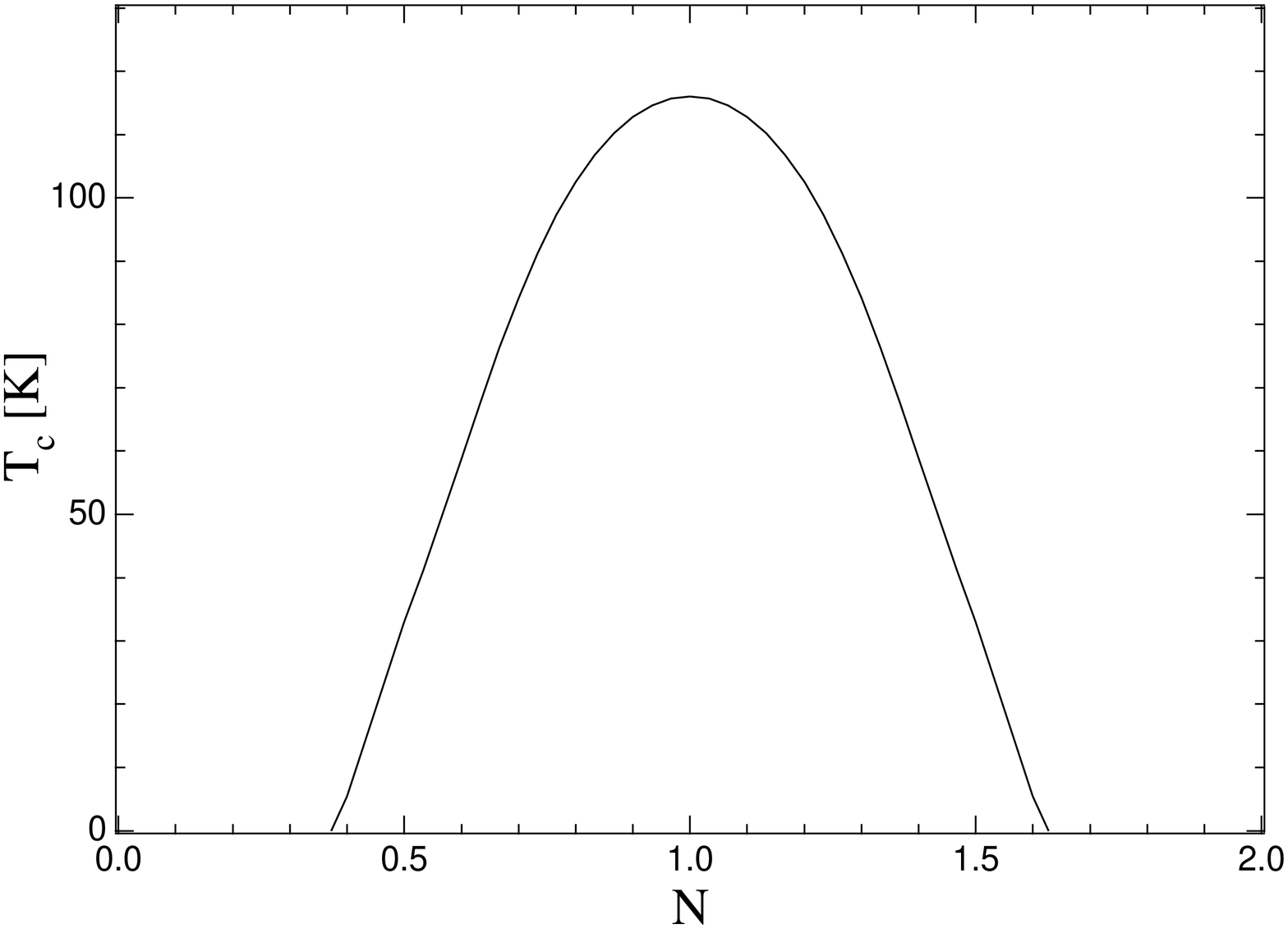}
\end{center}
\caption{$T_{ c } $ as a function of the hole concentration
for $ J_{ H } = 0.13 $ eV, $ J _{ K } = 0.2 $ eV and
$ t_{ pp } = 0.65 $ eV }
\label{FigTcxN}
\end{figure}
Fig. \ref{FigTcxN} shows our results for the
superconducting phase diagram
using
$ J_{ H } = 0.13 $ eV,
$ J_{ K } = 0.2 $ eV
and a first neighbors hopping dispersion relation between oxygen sites,
with $ t_{ pp } = 0.65 $ eV.
We obtain $ T_{ c } \sim 115 $ K as a maximum value for $ T_{ c } $
and our results are in good agreement with experiment.

\section*{Acknowledgement}
This work has been supported in part by FAPERJ and CNPq.

\end{document}